\documentclass[12pt,english]{article}
\usepackage[T1]{fontenc}
\usepackage[latin9]{inputenc}
\usepackage[letterpaper]{geometry}
\usepackage{textcomp}
\usepackage{url}
\usepackage{amsthm}
\usepackage{amsmath}
\usepackage{graphicx}
\usepackage{setspace}
\usepackage{amssymb}
\usepackage[authoryear]{natbib}
\makeatletter
\newcommand{\lyxaddress}[1]{
\par {\raggedright #1
\vspace{1.4em}
\noindent\par}
}
\theoremstyle{plain}
\newtheorem{thm}{Theorem}
 \theoremstyle{definition}
  \newtheorem{example}[thm]{Example}

\DeclareMathOperator{\unif}{U}

\DeclareMathOperator{\N}{N}

\DeclareMathOperator{\lfdr}{LFDR}

\makeatother

\usepackage{babel}

\begin{document}

\title{\textbf{\large ~}\\
Small-scale inference: Empirical Bayes and confidence methods
for as few as a single comparison\\
\textbf{~}}

\maketitle
~~\\
David R. Bickel

\lyxaddress{Ottawa Institute of Systems Biology\\
Department of Biochemistry, Microbiology, and Immunology\\
University of Ottawa; 451 Smyth Road; Ottawa, Ontario, K1H 8M5}
\begin{abstract}

By restricting the possible values of the proportion of null hypotheses
that are true, the local false discovery rate (LFDR) can be estimated
using as few as one comparison. The proportion of proteins with
equivalent abundance was estimated to be about 20\% for patient group
I and about 90\% for group II. The simultaneously-estimated LFDRs
give approximately the same inferences as individual-protein confidence
levels for group I but are much closer to individual-protein LFDR
estimates for group II. Simulations confirm that confidence-based
inference or LFDR-based inference performs markedly better for low
or high proportions of true null hypotheses, respectively.
\end{abstract}
\textbf{Keywords:} confidence distribution; empirical Bayes; Lindley's
paradox; local false discovery rate; multiple comparison procedure;
multiple testing; observed confidence level; restricted parameter
space

\section{\label{sec:Introduction}Introduction}

In the development of statistical methods for interpreting high-dimensional
genomics data, the challenges involved in analyzing genomics data
sets of much smaller scale have been largely overlooked, and yet such
data are routinely generated. Out of the thousands of genes in the
human genome, the expression levels of only on the order of 30 genes
are measured in a real-time polymerase chain reaction experiment.
Among the hundreds of thousands of proteins in the human proteome,
the abundance levels of only on the order of 200 proteins are measured
with mass spectrometry. The following idealization of the candidate-gene
approach to genetic association studies poses a problem encountered
in analyzing data from a small fraction of a large number of biological
features, with each feature corresponding to a different population
in the sampling theory sense. 
\begin{example}
\label{exa:normal}Consider $10^{6}$ populations such that $X_{i}\sim\N\left(\mu_{i},1\right)$
for $i=1,\dots,10^{6}$, where $\mu_{i}=2$ for $N_{1}$ values of
$i$ and $\mu_{i}=0$ for $10^{6}-N_{1}$ values of $i$. None of
the random values is observed except $x_{1}$, the realization of
$X_{1}$. The null hypothesis of interest is $\mu_{1}=0$. Let $\Phi$
and $\phi$ respectively denote the standard normal distribution function
and density function. Without any knowledge of $N_{1}$, few would
question the applicability of the p-value $1-\Phi\left(x_{1}\right)$.
On the other hand, in the absence of other information, the use of
$P\left(\mu_{1}=0;N_{1}\right)=1-N_{1}/10^{6}$ as an approximate,
nonsubjective prior probability of the null hypothesis in order to
obtain the approximate posterior probability\begin{equation}
P\left(\mu_{1}=0\vert x_{1};N_{1}\right)=\frac{\left(1-N_{1}/10^{6}\right)\phi\left(x_{1}\right)}{\left(1-N_{1}/10^{6}\right)\phi\left(x_{1}\right)+\left(N_{1}/10^{6}\right)\phi\left(x_{1}-2\right)}\label{eq:normal}\end{equation}
would not be controversial if $N_{1}$ were known. Suppose that $N_{1}$
is unknown but can be safely assumed to be between 1 and 100. Then,
for at least 99.99\% of the populations, the null hypothesis is true
and thus $1-\Phi\left(X_{1}\right)\sim\unif\left(0,1\right)$. By
contrast, for those same populations, $P\left(\mu_{1}=0\vert X_{1};\tilde{N}_{1}\right)\approx1$
with high probability regardless of the value $\tilde{N}_{1}$ between
1 and 100 that is guessed for $N_{1}$ in computing the posterior
probability. For instance, if $x_{1}=2$, then the p-value is $1-\Phi\left(2\right)=2.28\%$
even though the posterior probability of the null hypothesis is at
least $P\left(\mu_{1}=0\vert2;100\right)=99.93\%$ and possibly as
high as $P\left(\mu_{1}=0\vert2;1\right)=1-7.39\times10^{-6}$. \citet{ISI:A1957XF49000016}
thoroughly examined a similar {}``paradox'' from a more Bayesian
viewpoint.
\end{example}
The type of problem faced in Example \ref{exa:normal} will be attacked
by adapting methodology recently developed for gene expression microarray
data to two other settings: (1) those with data available for testing
only a much smaller number of hypotheses and (2) those with much smaller
proportions of null hypotheses that are true. 

Microarray technology enables the measurement of levels of gene expression
for thousands of genes in cells under two different conditions, conveniently
labeled as treatment and control. Which genes have differential expression
in the mean between the treatment and control populations? That large-scale
problem of multiple comparisons led \citet{RefWorks:53} to apply
the false discovery rate (FDR) of \citet{RefWorks:288} and to introduce
the local false discovery rate (LFDR). In accordance with its name,
the LFDR is a rate of Type I errors that would be incurred were the
null hypothesis rejected every time the same data are generated as
those actually observed. In the microarray context, the LFDR is
an empirical Bayes posterior probability of the null hypothesis that
a particular gene does not have differential expression, as in equation
\eqref{eq:normal}. More precisely, the LFDR is defined as the prior
probability of the null hypothesis conditional on the p-value or other
statistic that reduces the measured expression levels of the gene
to a single number \citep{efron_large-scale_2010}. 

Here, like in Example \ref{exa:normal}, the prior probability approximates
an unknown proportion of null hypotheses that are true, with each
null hypothesis corresponding to a different gene. In that sense,
the LFDR differs from a fully Bayesian posterior probability, which
requires the complete specification of the prior distribution of all
unknown parameters. Such specification usually involves prior probabilities
that correspond to hypothetical levels of belief rather than real
relative frequencies or proportions. Thus, whereas a purely Bayesian
prior is necessarily known in principle, empirical Bayes priors are
unknown.

Since the LFDR generally depends on parameters that do not have a
known prior distribution, the LFDR can only be estimated. Supposing,
however, that the LFDR could be known and neglecting any information
lost in reducing the data to a test statistic for each hypothesis,
Bayes decision rules based on the LFDR would have optimal Bayes risk.
That is, they would perform at least as well on average as any other
decision rule with respect to any bounded loss function. Knowledge
of the LFDR would require knowledge not only of the proportion of
null hypotheses that are true but also the distribution of the reduced
data under the alternative hypotheses. In that case, there would be
no objection against relying on the LFDR derived from Bayes's theorem
since frequentists by principle condition on the data in the presence
of a known population of parameter values (\citealp{RefWorks:985,Wilkinson1977,RefWorks:52};
\citealp{RefWorks:1459}; \citealp[p. 36]{AndersHald2007b}; \citet{Yuan20092458};
\citealp{fraser-2009}). With that knowledge, the unquestioned applicability
of the LFDR would hold regardless of the number of hypotheses that
correspond to measurements. As a result, the LFDR would apply to a
single comparison corresponding to a hypothesis randomly drawn from
the population (Example \ref{exa:normal}) no less than to multiple
comparisons spanning the entire population of hypotheses.

However, it is generally believed that the LFDR can only be adequately
estimated if there are data directly related to thousands of hypotheses.
For example, if data are only available for 20 genes, or, in the case
study of this paper, 20 proteins, then the LFDR is not considered
applicable. Indeed, empirical Bayes methods designed for several thousands of
comparisons do not necessarily work as well with smaller numbers of
hypotheses. 

In some respects, that limitation of the empirical Bayes framework
restricts the utility of multiple comparison procedures more generally.
The discussions of two empirical Bayes papers spanning the last three
decades \citep{Morris1983515,efron_correlated_2010} illustrate the
consensus that very different procedures seem suitable for different
numbers of comparisons. \citet{Westfall_efron_correlated_2010} emphasized
in his comment that whereas methods that control family-wise error
rates (FWERs) have insufficient statistical power for very large numbers
of comparisons, estimators of FDRs and LFDRs become unreliable for
small numbers of comparisons. \citet{rejoinder_efron_correlated_2010}
replied with a recommendation for FWER control for smaller numbers
of comparisons as a substitute for empirical Bayes estimation of the
FDR for larger numbers of comparisons. That conflicts with the viewpoint
of \citet{Morris1983b}, another pioneer of empirical Bayes procedures,
who resorted to fully Bayesian procedures for small numbers of comparisons.

The main purpose of this paper is to extend the scope of LFDR estimation
to the smallest possible scale: that of a single comparison.  The
investigation will involve modifying a successful method of LFDR estimation
and studying its relative performance in various contexts. It will
be compared to fully Bayesian inference under a default prior and
to the p-value interpreted inferentially with the aid of confidence
distributions. The importance of the p-value in the multiple comparison
framework lies in the fact that it is equal to the p-value adjusted
to control an error rate when only one comparison is made. For example,
with data for only a single hypothesis test, the achieved FDR, the
lowest value at which the FDR has guaranteed control, is equal to
the p-value \citep{RefWorks:288}.

Were such a method of small-scale LFDR estimation available for small-scale
genetic association studies, the widespread publication of significant
findings that could not be replicated \citep{morgenthaler_strategy_2007}
might have been avoided. The reason is that LFDR estimation takes
advantage of an estimate of the proportion of null hypotheses that
are true, which is crucial for extremely small proportions, whereas
p-values ignore that information, thereby inflating the Type I error
rate of testing a hypothesis picked at random. 
\begin{example}
\label{exa:genetic-association}

For testing hundreds of thousands of genetic variants for association
with disease,  FWER control in the tradition of Bonferroni, \citet{ISI:A19679545600024},
and \citet{ISI:A1979JY78700003} often, due to the large number of
tests, results in the rejection of few or no null hypotheses. The
alarming number of false positives found in candidate gene studies
\citep{morgenthaler_strategy_2007} at first seems to support such
adjustments of p-values for the number of tests in order to control
an FWER. However, the analogous history of false positives in candidate-gene
studies \citep{ISI:000171911000017}, in which much smaller numbers
of tests were performed in each study, shows that the number of tests
is not the source of the high false-positive rate. Rather, the root
of the problem lies more in the small number of disease-associated
variants compared to the total number of variants, irrespective of
how many happen to be measured. Thus, many join the \citet{RefWorks:199}
in questioning {}``the view that one should correct significance
levels for the number of tests performed to obtain `genome-wide significance
levels.''' In place of the number of tests performed, the \citet{RefWorks:199}
uses the proportion of variants that are associated with disease as
the prior probability of association, an approach that applies in
principle even to data representing only a single variant. That proportion
is thought to be between $10^{-6}$ and $10^{-4}$, as in Example
\ref{exa:normal}. 
\end{example}
Section \ref{sec:Empirical-Bayes-methods} introduces a parametric
method that enables empirical Bayes inference even in the absence
of multiple comparisons. Next, Section \ref{sec:Confidence-methods}
derives rival posterior distributions from confidence intervals under
fixed-parameter models. An application to proteomics data illustrates
the empirical Bayes and confidence methods in Section \ref{sec:Case-study}.
Section \ref{sec:Simulation-study} compares the performance of the
empirical Bayes and confidence methods for inference about a single
scalar parameter value that belongs to some population of parameter
values. The paper closes in Section \ref{sec:Discussion} with a discussion
of the resulting implications on whether empirical Bayes or confidence
strategies would be more suitable in a given context.

\section{\label{sec:Empirical-Bayes-methods}Empirical Bayes methods}

While methods of estimating the LFDR on the basis of nonparametric
density estimators clearly cannot apply to single-comparison data
\citep{efron_large-scale_2010}, it will be seen  that fully parametric
methods of LFDR estimation by maximum likelihood can do so under sufficiently
simple models. Since the empirical Bayes models that define the LFDR
have random parameters, the likelihood is not maximized over their
values but rather over the values of the hyperparameters specifying
the proportion of null hypotheses that are true and the distribution
of the reduced data under the alternative hypothesis. Such parameters,
if known, would entail knowledge of the LFDR (§\ref{sec:Introduction}).
More generally, the maximization of likelihood over hyperparameters
is called \emph{Type II maximum likelihood} as opposed to the Type
I maximum likelihood of models that lack random parameters \citep{GOOD01121966}.

\subsection{Hierarchical sampling model}

\subsubsection{\label{sub:Level-1}Level 1 of the model}

Consider a reference set of $\tilde{N}$ populations that includes
the $N$ populations sampled. Thus, $N$ is the number of comparisons
can be made on the basis of available data. For example, $\tilde{N}$
may be the number of genes in the genome, whereas $N$ is the number
of genes on the microarray that measures gene expression or is equal
to 1 if the expression of only a single gene is measured. Here, a
\emph{comparison} is understood as a hypothesis test or an effect-size
estimate.

Let $X_{i}$, an observable vector of dimension $n$, be a random
variable of a distribution $P_{\theta_{i},\lambda_{i}}$, which depends
on $\theta_{i}$, the parameter of interest, and on $\lambda_{i}$,
the nuisance parameter, for all $i\in\left\{ 1,\dots,\tilde{N}\right\} $.
Similarly, model $x_{j}$, the vector of $n$ observations, as a realization
of $X_{j}$ for all $j\in\left\{ 1,\dots,N\right\} $. 

Those data are reduced as follows. A \emph{random statistic} $U_{i}$
is a function of $X_{i}$, and an \emph{observed statistic} $u_{j}$
is a function of $x_{j}$, where the same function is applied to all
$i\in\left\{ 1,\dots,\tilde{N}\right\} $ and to all $j\in\left\{ 1,\dots,N\right\} $.
Thus, $u_{i}$ is a realization of $U_{i}$ for all $i\in\left\{ 1,\dots,N\right\} $. 

Supposing\emph{ }the distribution of \emph{$U_{i}$} is indexed by
the \emph{reduced parameter} $\delta_{i}$, a function of $\theta_{i}$
and $\lambda_{i}$, its probability mass function or density function
is denoted by $f\left(\bullet;\delta_{i}\right)$ for each $i\in\left\{ 1,\dots,\tilde{N}\right\} $.
It follows that the probability mass or density of $u_{i}$ is $f\left(u_{i};\delta_{i}\right)$
for all $i\in\left\{ 1,\dots,N\right\} $. Without loss of generality,
the $i$th null hypothesis is that $\theta_{i}=0$ or, equivalently,
$\delta_{i}=0$, for any $i\in\left\{ 1,\dots,\tilde{N}\right\} $.
\begin{example}
\label{exa:absTd}Suppose the expression level of each of $N$ genes
is measured for a total of $n^{\text{treat}}$ cell cultures treated
with a chemical and $n^{\text{control}}$ cell cultures not so treated.
The expression level of the $i$th gene is the logarithm of a measure
of the abundance of mRNA in the cells and is IID $\N\left(\theta_{i}^{\text{treat}},\lambda_{i}^{2}\right)$
within the treatment group and IID $\N\left(\theta_{i}^{\text{control}},\lambda_{i}^{2}\right)$
within the control group, $\lambda_{i}$ being the common standard
deviation. Then $T_{i}$, the equal-variance Student $t$ test statistic,
has a noncentral $t$ distribution with noncentrality parameter $\Delta_{i}=\left(\theta_{i}^{\text{treat}}-\theta_{i}^{\text{control}}\right)\left(1/n^{\text{treat}}+1/n^{\text{control}}\right)^{-1/2}/\lambda_{i}$
and $n-2=n^{\text{treat}}+n^{\text{control}}-2$ degrees of freedom;
this is abbreviated by $T_{i}\sim\text{Student}\left(\Delta_{i},n-2\right)$.
Then $U_{i}=\left|T_{i}\right|$ is very effective for inference about
$\delta_{i}=\left|\Delta_{i}\right|$. By implication, $U_{i}$ is
highly informative about the expression \emph{fold change} $\exp\left|\theta_{i}^{\text{treat}}-\theta_{i}^{\text{control}}\right|$,
the effect size most often estimated in reports of microarray data
analysis, and about whether $\theta_{i}^{\text{treat}}=\theta_{i}^{\text{control}}$
since that is necessary and sufficient for $\delta_{i}=0$. If $n^{\text{treat}}+n^{\text{control}}$
is large enough, then $T_{i}\,\dot{\sim}\,\N\left(\Delta_{i},1\right)$,
which entails that $U_{i}^{2}$ is approximately distributed as $\chi^{2}\left(\delta_{i}^{2},1\right)$,
the noncentral chi-square distribution with noncentrality parameter
$\delta_{i}^{2}$ and 1 degree of freedom. 
\end{example}
The most common model for analyzing genetic association data has the
same asymptotics.
\begin{example}
\label{exa:genetic-association-model}Example \ref{exa:genetic-association},
continued. In order to utilize genetic models such as the additive
model \citep{lewis_genetic_2002} and in order to account for effects
of covariates, genetic association data are typically analyzed using
the Wald approximation with logistic regression, yielding the statistic
$T_{i}$ equal to the (Type I) maximum likelihood estimate of the
log odds ratio divided by the estimated standard error of that estimate
for variant $i$ of $N$. The statistic $U_{i}=\left|T_{i}\right|$
is highly informative about the absolute value of the log odds ratio
and whether it is equal to 0, as under the null hypothesis of no association
between the genotype and the trait. For a sufficiently high number
of case and control subjects, $U_{i}^{2}\,\dot{\sim}\,\chi^{2}\left(\delta_{i}^{2},1\right)$,
as in Example \ref{exa:absTd}. 
\end{example}

\subsubsection{\label{sub:Level-2}Level 2 of the model}

The first level of the hierarchical model describes the variability
of the expression levels of each gene or other population that corresponds
to a comparison (§\ref{sub:Level-1}). To represent variability between
populations or comparisons, $\delta_{i}$ is now modeled as the random
variable equal to 0 with probability $\pi_{0}$, equal to some $\delta^{\left(1\right)}\ne0$
with probability $\pi_{1}$, equal to some $\delta^{\left(2\right)}\notin\left\{ 0,\delta^{\left(1\right)}\right\} $
with probability $\pi_{2}$, ..., and equal to some $\delta^{\left(K\right)}\notin\left\{ 0,\delta^{\left(K\right)}\right\} $
with probability for a $K\in\left\{ 1,2,\dots\right\} $. The alternative-hypothesis
parameters constitute $\psi,$ a matrix with $\left\langle \pi_{1},\dots,\pi_{K}\right\rangle $
and $\left\langle \delta^{\left(1\right)},\dots,\delta^{\left(K\right)}\right\rangle $
as its two columns. 

Then the unknown hyperparameters are $\pi_{0}$ and $\psi$, and the
probability mass function or density function of $X_{i}$ is the finite
mixture \[
\bar{f}\left(\bullet;\pi_{0},\psi\right)=\pi_{0}f\left(\bullet;0\right)+\sum_{k=1}^{K}\pi_{k}f\left(\bullet;\delta^{\left(k\right)}\right)\]
for all $i\in\left\{ 1,\dots,\tilde{N}\right\} $. The random indicator
$\nu_{i}$ will determine whether the null hypothesis is true $\left(\nu_{i}=1\right)$
or false $\left(\nu_{i}=0\right)$ for all $i\in\left\{ 1,\dots,\tilde{N}\right\} $.
It is assumed that $\tilde{N}$ is large enough that $P\left(\nu_{i}=1\right)=\pi_{0}$
is approximately $\sum_{i=1}^{\tilde{N}}\nu_{i}/\tilde{N}$, the proportion
of null hypotheses that are true. 

The \emph{local false discovery rate}, $P\left(\nu_{i}=1\vert U_{i}=u_{i}\right)$
by definition, is \[
\lfdr\left(u_{i};\pi_{0},\psi\right)=\frac{P\left(\nu_{i}=1\right)\bar{f}\left(u_{i}\vert\nu_{i}=1;\pi_{0},\psi\right)}{\bar{f}\left(u_{i};\pi_{0},\psi\right)}=\frac{\pi_{0}f\left(u_{i};0\right)}{\bar{f}\left(u_{i};\pi_{0},\psi\right)}\]
by Bayes's theorem. As this LFDR is unknown only because $\pi_{0}$
and $\psi$ are unknown, it may be estimated by Type II maximum likelihood,
as will now be seen.

\subsection{\label{sub:Type-II-MLE}Type II maximum likelihood}

The hyperparameters are estimated by $\hat{\pi}_{0}$ and $\hat{\psi}$,
the values of $\pi_{0}$ and $\psi$ at which the likelihood\[
\prod_{i=1}^{N}f\left(x_{i};\pi_{0},\psi\right)\]
attains its maximum subject to the constraints that $\sum_{k=1}^{K}\pi_{k}=1$
and $0\le\pi_{k}\le1$. Then $\lfdr\left(u_{i};\hat{\pi}_{0},\hat{\psi}\right)$
is the maximum likelihood estimate of the LFDR. \citet{Pawitan20054435},
\citet{ParametricMixtureLFDR}, and \citet{GWAselect} employed this
method of estimating the LFDR under fully parametric finite mixtures.

To prevent overfitting in the form of excessive variance in the estimates,
the value of $K$ must be smaller for smaller values of $N$. For that reason, \citet{mediumScale} suggested $K=1$ when $N<1000$. That
model is simpler than those of higher values of $K$: the only free
parameters are $\pi_{0}$, the approximate proportion of null hypotheses
that are true, and $\delta^{\left(1\right)}$, the value of the reduced
parameter indexing the alternative distribution. However, it is not
simple enough for a single comparison $\left(N=1\right)$, for in
that case, $\hat{\pi}_{0}=0$ almost always. 

More generally, whenever $N$ is deemed too small for reliable estimation
of $\hat{\pi}_{0}$ with $\pi_{0}$ only restricted to the interval
$\left[0,1\right]$, it will be further constrained to the strictly
smaller interval $\left[\pi_{0}^{-},\pi_{0}^{+}\right]$, a proper
subset of $\left[0,1\right]$ with the specified bounds $\pi_{0}^{-}$
and $\pi_{0}^{+}$ such that $0\le\pi_{0}^{-}\le\pi_{0}^{+}\le1$.
Thus, the proposed method guarantees that $\pi_{0}^{-}\le\hat{\pi}_{0}\le\pi_{0}^{+}$
even for the lowest values of $N$. 

In the case of $N=1$, there is overfitting in the sense that $\hat{\pi}_{0}=\pi_{0}^{-}$
almost always. Likewise, for small values of $N$, $\hat{\psi}$ is
not an optimal estimator of $\psi$. Thus, improvements such as those
based on predictive distributions are certainly possible \citep[e.g.,][]{NMWL}.
Nonetheless, the application (§\ref{sec:Case-study}) and simulations
(§\ref{sec:Simulation-study}) demonstrate that even the simple method
introduced here can perform substantially better than methods that
take no account of the hierarchical structure of the data. It will
be seen that with certain distributions of unknown parameter values,
even extremely crude estimates of the hyperparameters are preferable
to no estimates at all.

To prevent problems with numerically maximizing the likelihood, the
reduced parameter $\delta^{\left(1\right)}$ was constrained under
the alternative hypothesis to have a lower bound of $10^{-3}$ for
Sections \ref{sec:Case-study} and \ref{sec:Simulation-study}, but
none of the results was sensitive to the value of that bound.
\section{\label{sec:Confidence-methods}Confidence methods}

This section confines attention to the single-level model consisting
of the model of Section \ref{sub:Level-1} with fixed parameters rather
than the random parameters of Section \ref{sub:Level-2}. The concept
of confidence posterior distributions will be reviewed to set the
stage for the observed confidence levels to consider as viable alternatives
to LFDRs.

Let $\Theta\subseteq\mathbb{R}^{1}$ denote the \emph{parameter space}
of each fixed parameter value $\theta_{i}$ in the sense that it is
the smallest set in which $\theta_{i}$ is known to lie. Likewise,
let $\Lambda$ denote the parameter space of each $\lambda_{i}$.
Whereas the nuisance parameter $\lambda_{i}$ may be a scalar or vector,
it is assumed that the interest parameter $\theta_{i}$ is a scalar,
i.e., that $\Theta\subseteq\mathbb{R}^{1}$.

Consider $\vartheta_{i}$, the random variable that has probability
distribution $P\left(\bullet;u_{i}\right)$ on $\Theta$ such that\begin{equation}
P\left(\vartheta_{i}\le\theta_{i};u_{i}\right)=P_{\theta_{i},\lambda_{i}}\left(U_{i}\ge u_{i}\right)\label{eq:confidence-posterior}\end{equation}
for all $\theta_{i}\in\Theta$ and $\lambda_{i}\in\Lambda$, where
$U_{i}$ is a scalar statistic determined by $P_{\theta_{i},\lambda_{i}}$,
the sampling distribution of $X_{i}$ introduced in Section \ref{sub:Level-1}.
The random elements of the equation are $\vartheta_{i}$ on the left-hand
side but $U_{i}$ on the right-hand side. 

The probability measure $P\left(\bullet;u_{i}\right)$ is the \emph{confidence
\emph{posterior }distribution} of $\theta_{i}$. The word \emph{confidence
}emphasizes the property that the interval bounded by the $\beta_{1}$-quantile
and the $\beta_{2}$-quantile of $\vartheta_{i}$ is a $\left(\beta_{2}-\beta_{1}\right)100\%$
confidence interval in the sense that it has a $\left(\beta_{2}-\beta_{1}\right)100\%$
frequentist probability of including $\theta_{i}$ \citep{Efron19933,RefWorks:127,RefWorks:130}.
While the term \emph{posterior} correctly indicates the dependence
of the parameter distribution $P\left(\bullet;u_{i}\right)$ on the
observed statistic $u_{i}$ \citep{conditional2009,CoherentFrequentism},
it is not necessarily a Bayesian posterior, a conditional prior distribution
given $U_{i}=u_{i}$. For example, $P\left(\theta^{-}\le\vartheta_{i}\le\theta^{+};u_{i}\right)$
is the confidence posterior probability of the hypothesis that the
parameter of interest lies between the fixed values $\theta^{-}$
and $\theta^{+}$ and yet need not correspond to any Bayesian posterior
probability of the hypothesis. \citet{Polansky2007b} calls $P\left(\theta^{-}\le\vartheta_{i}\le\theta^{+};u_{i}\right)$
the \emph{observed confidence level} of the hypothesis; cf. \citet{RefWorks:249}. 
\begin{example}
\label{exa:Td}Example \ref{exa:absTd}, continued. For simplicity,
the statistic is changed to $U_{i}=T_{i}$, which is useful for inference
about the value of $\theta_{i}=\theta_{i}^{\text{treat}}-\theta_{i}^{\text{control}}$.
Since $T_{i}\sim\text{Student}\left(\Delta_{i},n-2\right)$, equation
\eqref{eq:confidence-posterior} implies that $\vartheta_{i}/\hat{\sigma}_{i}\sim\text{Student}\left(t_{i},n-2\right)$,
where $\hat{\sigma}_{i}$ is the typical pooled estimate of the standard
error of the sample mean difference between treatment and control
\citep{RefWorks:127}. Thus, the confidence posterior distribution
of the parameter of interest is equivalent to the Bayesian posterior
distribution resulting from the improper priors according to which
the mean and the logarithm of the standard deviation are uniform on
the real line. Coherence in the Bayesian sense would then require
that the same posterior distribution be used for inference about $\left|\theta_{i}\right|$,
e.g., \begin{eqnarray}
P\left(\left|\vartheta_{i}\right|=0;t_{i}\right)=P\left(\vartheta_{i}=0;t_{i}\right) & = & P\left(\vartheta_{i}\le0;t_{i}\right)-\lim_{\epsilon\rightarrow0+}P\left(\vartheta_{i}\le0-\epsilon;t_{i}\right)\nonumber \\
 & = & P_{0,\lambda_{i}}\left(U_{i}\ge u_{i}\right)-\lim_{\epsilon\rightarrow0+}P_{0+\epsilon,\lambda_{i}}\left(U_{i}\ge u_{i}\right)=0.\label{eq:zero-confidence}\end{eqnarray}
For $\lambda_{i}=1$ and large $n$, $\vartheta_{i}^{2}\,\dot{\sim}\,\chi^{2}\left(\delta_{i}^{2},1\right)$,
which \citet{RefWorks:1431} presented as the fiducial distribution
for inference about $\theta_{i}^{2}$, contrasting its interval estimates
with confidence intervals.
\end{example}
The next example extracts a different confidence posterior distribution
from the same statistical model.
\begin{example}
\label{exa:Stein-Wilkinson}Example \ref{exa:absTd}, continued. Let
$U_{i}=\left|T_{i}\right|$ to draw inferences about $\theta_{i}=\left|\theta_{i}^{\text{treat}}-\theta_{i}^{\text{control}}\right|$.
By equation \eqref{eq:confidence-posterior}, $P\left(\bullet;u_{i}\right)$,
the confidence posterior distribution of $\vartheta_{i}$, is defined
by \[
P\left(\vartheta_{i}\le\theta_{i};u_{i}\right)=P_{\theta_{i},\lambda_{i}}\left(\left|T_{i}\right|\ge u_{i}\right).\]
Because $T_{i}\sim\text{Student}\left(0,n-2\right)$ under the null
hypothesis that $\theta_{i}=0$, the confidence posterior probability
that the null hypothesis is true is equal to the usual two-sided p-value:
\begin{equation}
P\left(\vartheta_{i}=0;u_{i}\right)=P\left(\vartheta_{i}\le0;u_{i}\right)=P_{0,\lambda_{i}}\left(\left|T_{i}\right|\ge u_{i}\right).\label{eq:confidence-shrinkage}\end{equation}
This is a clear counterexample to the observation of \citet{Polansky2007b}
and \citet{conditional2009} that many confidence posteriors e.g.,
that of Example \ref{exa:Td}, put no probability mass on any simple
hypothesis.
\end{example}

Like the Bayesian posterior, the confidence posterior can be used
to make coherent decisions given a loss function \citep{conditional2009,CoherentFrequentism}.
In the metaphor of an intelligent agent, whereas the Bayesian posterior
describes the decisions made by an agent committed to a particular
prior distribution, the confidence posterior describes the decisions
made by an agent that interprets confidence levels from a particular
procedure as levels of certainty \citep{frequentistReasoning}. Thus,
the confidence posterior enables direct performance comparisons
between frequentist procedures and Bayesian and empirical Bayes posteriors,
as will be seen in Sections \ref{sec:Case-study} and \ref{sub:Hypothesis-testing}.

\section{\label{sec:Case-study}Application to proteomics data}

Alex Miron\textquoteright{}s lab at the Dana-Farber Cancer Institute
recorded the abundance level of each of 20 plasma proteins for every
woman of two breast-cancer groups (55 HER2-positive women and 35 mostly-ER/PR-positive
women) and of a control group (64 healthy women) \citep{ProData2009b}.
After adding the 25th percentile of the abundance levels within the
control group to all abundance levels in order to ensure that the
adjusted levels were positive \citep{mediumScale}, the logarithms
of the adjusted levels of a given gene were modeled as quantities
drawn from a normal distribution with the same variance.

In comparing each breast-cancer group to the control group, the data
for each protein were reduced to the absolute value of the equal-variance
\emph{t}-statistic, which has a Student \emph{t} distribution under
the null hypothesis of no difference between groups and a noncentral
Student \emph{t} distribution with noncentrality parameter $\delta$
under the alternative hypothesis of a nonzero mean difference, as
in Example \ref{exa:absTd}. 

In order to analyze the data of all proteins simultaneously, it was
assumed that the reduced data of all proteins with differential abundance
levels are absolute values of variates drawn from the same noncentral
\emph{t} distribution, the noncentrality parameter of which is denoted
by $\delta$. The assumption enabled computing $\hat{\pi}_{0}$ and
$\hat{\delta}$, the maximum likelihood estimates of $\pi_{0}$ and
$\delta$, using the empirical Bayes method of Section \ref{sub:Type-II-MLE}
with the constraint that $0\%\le\pi_{0}\le100\%$. For comparison,
the data of each protein were then analyzed individually by using
the confidence and empirical Bayes methods as if it were the only
protein with measured expression. 

\begin{figure}
\includegraphics[scale=0.8]{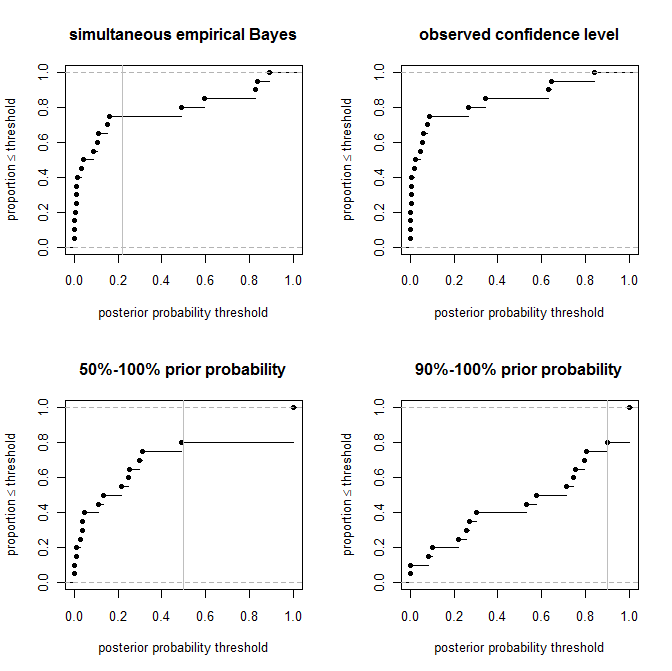}

\caption{Empirical distribution functions of the posterior probability that
a given protein has equivalent abundance between the HER2-positive
and control groups. The four methods compared are described in the
text.\label{fig:postAB}}

\end{figure}

\begin{figure}
\includegraphics[scale=0.8]{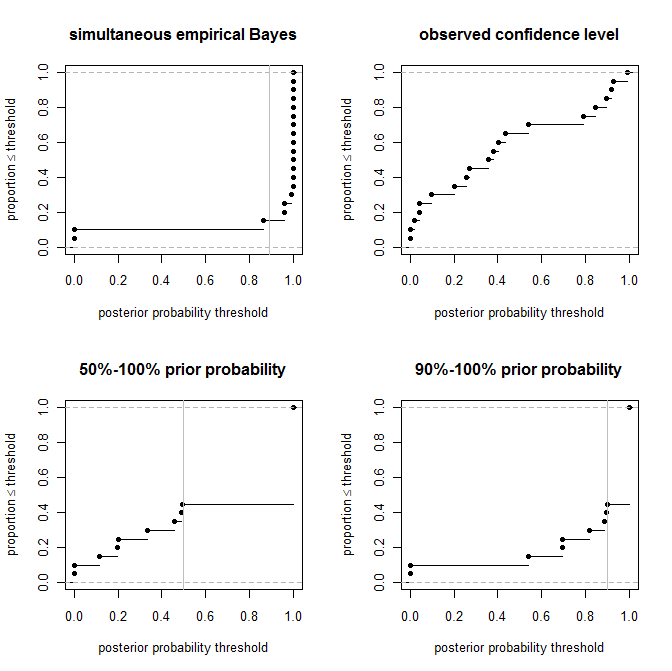}

\caption{Empirical distribution functions of the posterior probability that
a given protein has equivalent abundance between the ER/PR-positive
and control groups. Each plot corresponds to a method described in
the text.\label{fig:postCB}}

\end{figure}

The results are summarized in Figures \ref{fig:postAB} and \ref{fig:postCB}.
Within each figure, the posterior probability estimates of the top-left
plot are the LFDRs estimated by substituting $\hat{\pi}_{0}$ and
$\hat{\delta}$ for $\pi_{0}$ and $\delta$, with the vertical line
specifying the value of $\hat{\pi}_{0}$. Each posterior probability
of each top-right plot is the observed confidence level of the null
hypothesis of equivalent abundance between cancer and control groups
as recorded by equation \eqref{eq:confidence-shrinkage}. The bottom
two plots of each figure report the LFDRs estimated separately for
each protein by maximizing the likelihood with the constraints that
$\pi_{0}\ge50\%$ (bottom-left plot) and $\pi_{0}\ge90\%$ (bottom-left
plot), with the vertical lines drawn at 50\% and 90\%, respectively.

Since only the top-left plot of each figure represents the simultaneous
use of the data for all proteins, it serves as the reference for evaluating
the three methods of analyzing the data of each protein in isolation
from the other data. As seen in Figure \ref{fig:postAB}, the observed
confidence levels closely match the simultaneously estimated LFDRs
for the HER2-control group. By contrast, the individual-protein LFDR
estimates come much closer than the observed confidence levels to
the simultaneously estimated LFDRs for the ER/PR-control group (Figure
\ref{fig:postCB}). The explanation for that difference between comparisons
is that the estimated proportion of equivalent-abundance proteins
is low for the first group $\left(\hat{\pi}_{0}\doteq22\%\right)$
but high for the second group $\left(\hat{\pi}_{0}\doteq89\%\right)$. 

\section{\label{sec:Simulation-study}Simulation studies}

The simulation studies of the following subsections were carried out
in the scenario of $T_{i}\sim\N\left(\Delta_{i},1\right)$, $U_{i}\sim\left|T_{i}\right|$,
and $\delta_{i}=\left|\Delta_{i}\right|$ since it represents the
asymptotics of a wide variety of situations encountered in practice,
including those of protein abundance (§\ref{sec:Case-study}), gene
expression (Example \ref{exa:absTd}), and genetic association (Example
\ref{exa:genetic-association-model}). Specifically, the test statistics
were the absolute values of the realizations drawn from the normal
distribution with mean $\delta=0$ and variance 1 under the null hypothesis
and from the normal distribution with mean $\delta\in\left\{ 2,4\right\} $
and variance 1 under the alternative hypothesis. The mean error in
estimating the truth of the null hypothesis (§\ref{sub:Hypothesis-testing})
or the rate at which interval estimates cover $\delta$ (§\ref{sub:Effect-size-estimation})
then approximated the expected error or coverage probability of each
single-comparison method under the null and alternative hypotheses. 

Such approximations enabled approximating the expected error and coverage
probability for any proportion $\pi_{1}$ of null hypotheses that
are false as the weighted average of the expected error or coverage
probability with weight $1-\pi_{1}$ for the null hypothesis and $\pi_{1}$
for the alternative hypothesis. This quantifies the average performance
of applying each single-comparison method to data drawn from a randomly
selected hypothesis.

\subsection{\label{sub:Hypothesis-testing}Hypothesis testing}

The posterior probability that a method attributes to the null hypothesis
is its estimate of the value of the indicator $\nu_{i}$ that equals
1 if the null hypothesis is true or 0 if not (§\ref{sub:Level-2}).
Each method's estimation performance is here defined in terms of the
mean squared error (expected quadratic loss) for two reasons. First,
admissibility under quadratic loss is necessary and sufficient for
certain desirable properties relevant to conditional inference \citep{RefWorks:1081}.
Second, quadratic loss is the only proper scoring rule for probabilities
that (a) depends only on the difference between the estimator and
estimand and (b) remains unchanged if the estimator and estimand trade
places \citep{Savage1971b}. The square root of the expected quadratic
loss is easily interpreted as an average estimation error.

The present adoption of the confidence posterior probability of
equation \eqref{eq:confidence-shrinkage} is equivalent to interpreting
the p-value as an estimate of the indicator of whether the null hypothesis
is true. The p-value used this way does not require a significance
threshold and can dominate estimates defined to equal 0 if the p-value
is below such a threshold and equal to 1 otherwise \citep{Hwang1992490}.
Fixed-probability tails will be more appropriate for constructing
the confidence intervals of Section \ref{sub:Effect-size-estimation}
since it, unlike the present section, in effect imposes a 0-1 loss
function \citep{RefWorks:1081}.

\begin{figure}
\includegraphics[scale=0.5]{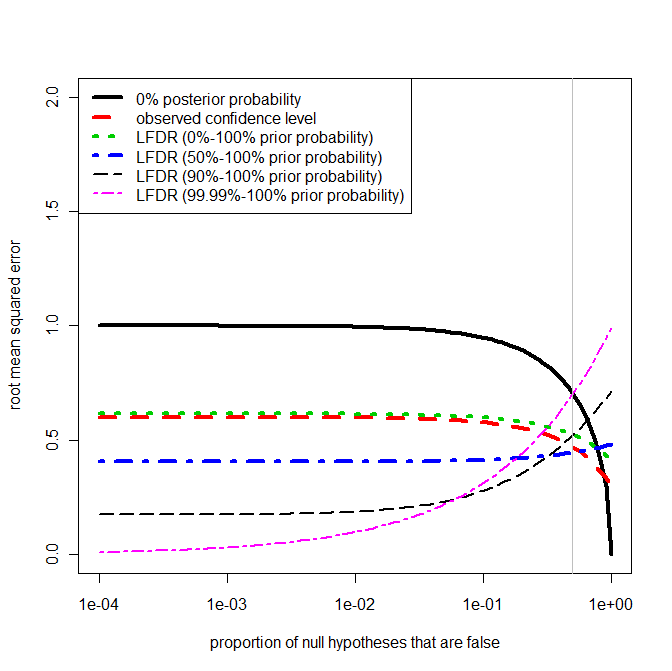}

\caption{Square root of the mean quadratic error of the (estimated) posterior
probabilities of null hypothesis truth versus $\pi_{1}=1-\pi_{0}$.
Reduced data were simulated from the unit-variance normal distributions
of means 0 (true null hypothesis) and 2 (false null hypothesis).\label{fig:rmse2}}

\end{figure}

\begin{figure}
\includegraphics[scale=0.5]{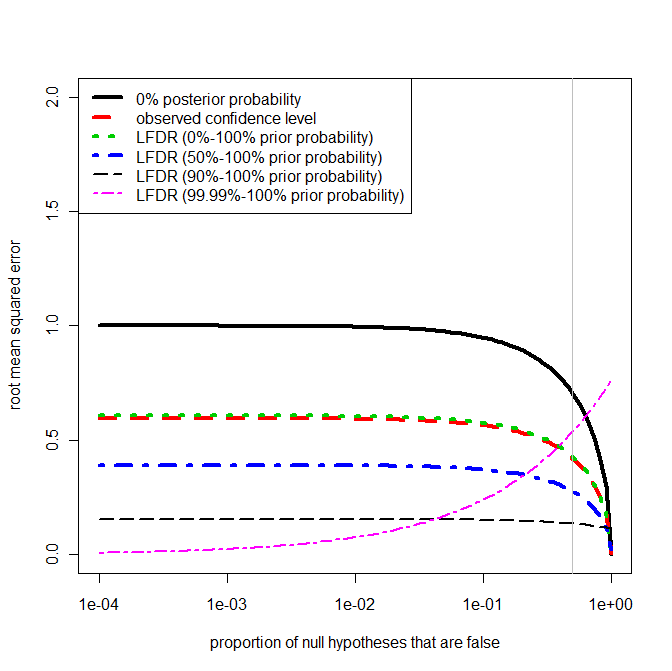}

\caption{Square root of the mean quadratic error of the (estimated) posterior
probabilities of null hypothesis truth versus $\pi_{1}=1-\pi_{0}$.
Reduced data were simulated from the unit-variance normal distributions
of means 0 (true null hypothesis) and 4 (false null hypothesis).\label{fig:rmse4}}

\end{figure}

On the basis of 100 realizations of the statistic drawn from each
of the three normal distributions $\N\left(0,1\right)$, $\N\left(2,1\right)$,
and $\N\left(4,1\right)$, Figures \ref{fig:rmse2} and \ref{fig:rmse4}
compare the mean quadratic loss of several methods of hypothesis testing
in the general form of assigning posterior probability to the null
hypothesis. The vertical lines are drawn at $\pi_{1}=50\%$. The \emph{0\%
posterior probability }represents any method that necessarily assigns
no probability mass to the simple null hypothesis, including improper-prior
Bayesian updating and all other methods yielding posterior density
functions (Example \ref{exa:Td}). The \emph{observed confidence level}
is the confidence posterior probability given by equation \eqref{eq:confidence-shrinkage}
with infinite degrees of freedom. Each of the four methods of estimating
the LFDR imposes a different constraint on $\pi_{0}$ when maximizing
the likelihood.%

\subsection{\label{sub:Effect-size-estimation}Effect-size estimation}

An interval estimate of the effect size $\left|\delta\right|$ is
the interval between two quantiles of a posterior distribution of
$\left|\delta\right|$, whether a confidence posterior, a Bayesian
posterior, or an empirical Bayes posterior. For example, the central
or equal-tail $\left(1-\alpha\right)100\%$ confidence interval corresponding
to a confidence posterior is the interval between its $\alpha/2$
and $1-\alpha/2$ quantiles. The coverage rate of an interval estimate
is its probability of including the true value of the interest parameter,
$\left|\delta\right|$ in the case of the simulation studies. 

\begin{figure}
\includegraphics[scale=0.5]{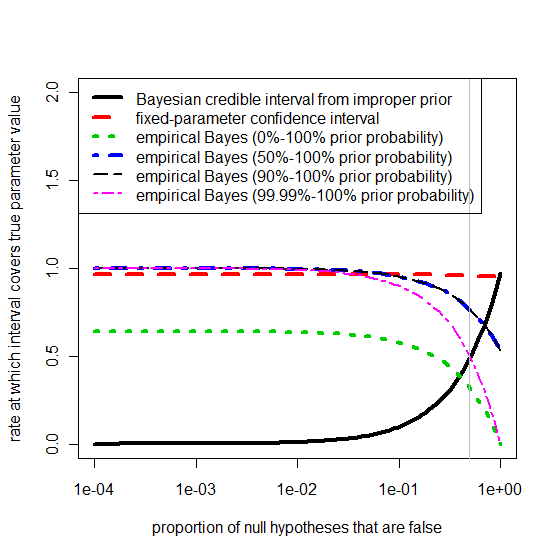}

\caption{Proportion of 95\% interval estimates that include the true value
of the mean versus $\pi_{1}=1-\pi_{0}$. Reduced data were simulated
from the unit-variance normal distributions of means 0 (true null
hypothesis) and 2 (false null hypothesis). The 50-100\% and 90\%-100\%
curves coincide.\label{fig:cover2}}

\end{figure}

Figure \ref{fig:cover2} displays the coverage rates of the equal-tail
95\% interval estimates for simulating 800 observed test statistics
from the null distribution and another 800 from the alternative distribution
with $\delta=2$. The displayed coverage rates are visually indistinguishable
from those instead using 800 draws from the $\delta=4$ distribution. 

The six posterior distributions of Figure \ref{fig:cover2} are those
of Section \ref{sub:Hypothesis-testing}, again with the vertical
line at $\pi_{1}=50\%$. The improper Bayesian posterior induced by
the uniform prior distribution of $\delta$ represents the class of
0\%-posterior methods (Example \ref{exa:Td}). Its interval estimates
were criticized by \citet{RefWorks:1431} and \citet{Wilkinson1977}
in favor of the confidence intervals of Figure \ref{fig:cover2}.
Its assignment of 0\% posterior probability to the null hypothesis
is evident from equation \eqref{eq:zero-confidence}. 

\section{\label{sec:Discussion}Discussion and conclusions}

The proposed method of constraining $\pi_{0}$ requires no more subjective
input than the popular methods of estimating the LFDR that rely on
nonparametric density estimation: they depend on the assumption that
$\pi_{0}$ be greater than about 90\% \citep{RefWorks:55}. With sufficiently
high choices of $\pi_{0}$, all such methods tend to be conservative.

The objection may be raised that all such choices are unnecessary
given the guaranteed coverage rates of fixed-parameter confidence
intervals. Indeed, although Bayesian and empirical Bayes methods can
cover the true parameter at slightly higher rates, they can also have
much worse coverage than confidence intervals. For example, empirical
Bayes intervals based on LFDR estimation have poor coverage at high
values of $\pi_{1}$ (Figure \ref{fig:cover2}). 

However, the main advantage of LFDR-based interval estimates over
fixed-parameter confidence intervals lies not in the potential increase
in the coverage rate but rather in the striking reduction in their
width \citep{RefWorks:1273,efron_large-scale_2010,hierarchy}. That
is especially true for lower values of $\pi_{1}$, as can be seen
from the greater and greater concentration of posterior probability
mass at the null hypothesis as $\pi_{1}\rightarrow0$ (Figures \ref{fig:rmse2}
and \ref{fig:rmse4}). Whenever the posterior probability of the null
hypothesis is at least 97.5\%, which happens with close to 100\% frequency
for high values of the lower bound $\pi_{0}^{-}$, the 95\% interval
estimate is $\left[0,0\right]$. That interval has zero width and
yet will cover the true value at a rate of $1-\pi_{1}$, the proportion
of null hypotheses $\left(\theta_{i}=0\right)$ that are true.

The value of $\pi_{1}$ also determines whether the LFDR approach
performs better or worse than the confidence approach in the context
of inferring whether or not a null hypothesis is true. For $\pi_{1}\dot{\le}10\%$,
there is substantial improvement in inference even when $\pi_{0}^{-}$
is far from $1-\pi_{1}$ (Figures \ref{fig:postCB}, \ref{fig:rmse2},
and \ref{fig:rmse4}). 

Among others, \citet{Lindley1957b} and \citet{RefWorks:1007} contrasted
Bayesian posterior probabilities of simple null hypotheses with p-values
before the LFDR was defined. The results of \citet{RefWorks:1007}
hold without their reliance on the misinterpretation of the p-value
as a Bayesian posterior probability since, in confidence-posterior decision theory \citep{conditional2009,CoherentFrequentism},
the two-sided p-value can be a legitimate confidence posterior probability
\eqref{eq:confidence-shrinkage}. \citet{RefWorks:1007} found that
the p-value can be far from the actual error rate, which necessarily
depends on $\pi_{1}$, the proportion of null hypotheses that are
false, whether or not that proportion is known. That, however, is
insufficient for concluding that Bayesian testing is superior: in
low-information situations, Bayesian posterior probabilities will
also be far from those that would be computed with knowledge of $\pi_{1}$
and other model parameters. For the practical scientist who does not
want to know about error rates but instead whether or not the null
hypothesis is true, the more important criterion is whether Bayesian
posterior probabilities or p-values come closer to $\nu_{i}$, the
indicator of the truth of the $i$th null hypothesis.

Using that criterion actually favored the p-value as an observed confidence
level over the empirical Bayes methods for $\pi_{1}\dot{\ge}50\%$
(Figures \ref{fig:postAB}, \ref{fig:rmse2}, and \ref{fig:rmse4}).
That largely vindicates the use of confidence-based methods when all
that is known about the parameter of interest is encoded either in
the model or in the test appropriate for a plausible null hypothesis
(§\ref{sec:Confidence-methods}). 

Nonetheless, even with the vague information that the hypothesis
tested belongs to a relevant class in which most null hypotheses are
true, rough guesses at $\pi_{0}^{-}$ can bring notable improvements
in inference accuracy. An extreme case is that of genetic association
studies (Example \ref{exa:genetic-association}), for which $\pi_{1}^{-}=10^{-6}$
and $\pi_{1}^{+}=10^{-4}$ are reasonable lower and upper bounds of
the proportion of SNPs associated with a given disease \citep{RefWorks:199}. 

The need to consider $\pi_{1}$ when making statistical inferences
cannot be avoided by running algorithms that automatically control
the FDR or FWER. The fundamental difference between the LFDR and the
FDR is exposed at lower numbers of comparisons and especially at the
single-comparison scale. Since FDR control reduces to standard hypothesis
testing when there is only a single test \citep{RefWorks:288}, the
achieved FDR, like any achieved FWER, is the unadjusted p-value and
thus is suitable in the same high-$\pi_{1}$ situations.

\section*{Acknowledgments}

The \texttt{Biobase} \citep{RefWorks:161} package of \texttt{R} \citep{R2008}
facilitated the computations.This research was partially supported  by the Canada Foundation for
Innovation, by the Ministry of Research and Innovation of Ontario,
and by the Faculty of Medicine of the University of Ottawa. 

\begin{flushleft}
\bibliographystyle{elsarticle-harv}
\bibliography{refman}

\par\end{flushleft}

\newpage{}

\end{document}